\documentclass{sbrt2019port}

\usepackage{url}
\usepackage{csquotes}

\begin{document}

\title{Ray-Tracing 5G Channels from Scenarios with Mobility Control of Vehicles and Pedestrians}

\author{Ailton Oliveira, Marcus Dias, Isabela Trindade, Aldebaro Klautau
\thanks{Ailton Oliveira, Marcus Dias, Isabela Trindade and Aldebaro Klautau  are with LASSE - Telecommunications, Automation and Electronics Research and Development Center, Bel\'em-PA, Brazil. E-mails: {ailton.pinto@itec.ufpa.br, marcus.dias@itec.ufpa.br, isabela.trindade@itec.ufpa.br, aldebaro@ufpa.br. Thanks to CNPq and CAPES for financial aid.}}}


\maketitle

\markboth{XXXVII SIMP{\'O}SIO BRASILEIRO DE TELECOMUNICAC{\~O}ES E PROCESSAMENTO DE SINAIS - SBrT2019, 29/09/2019--02/10/2019, PETR{\'O}POLIS, RJ}{XXXVII SIMPÓSIO BRASILEIRO DE TELECOMUNICAC{\~O}ES E PROCESSAMENTO DE SINAIS - SBrT2019, 29/09/2019--02/10/2019, PETR{\'O}POLIS, RJ} 

\begin{abstract}
Millimeter waves is one of 5G networks strategies to achieve high bit rates. Measurement campaigns with these signals are difficult and require expensive equipment. In order to generate realistic data this paper refines a methodology for \enquote{virtual} measurements of 5G channels, which combines a simulation of urban mobility with a ray-tracing simulator. The urban mobility simulator is responsible for controlling mobility, positioning pedestrians and vehicles throughout each scene while the ray-tracing simulator is repeatedly invoked, simulating the interactions between receivers and transmitters. The orchestration among both simulators is done using a Python software. To check how the realism can influence the computational cost, it was made a numerical analyse between the number of faces and the simulation time.
\end{abstract}

\begin{keywords}
	5G, millimeter waves, mobility, 3D models.
\end{keywords}

\section{Introduction}

Research and development of lower 5G layers deal with a relatively limited amount of data. For example, millimeter waves (mmWave) measurements for researches in 5G multiple-input multiple-output (MIMO) demand very expensive equipment and, possibly, measurement campaigns in elaborated external areas~\cite{maccartney_flexible_2017}.

In the current context, generating propagation channel data is a reasonable way to alleviate data scarcity while benefiting from an accuracy associated with ray-tracing (RT). For example, RT can handle 5G requirements such as spatial consistency, which has been a challenge to stochastic modeling~\cite{rumney_critical_2017}.

The lack of freely available data undermines evaluations and data-driven researches. Some researchers prefer use simplified models in their RT simulations, using cubes to represent the objects in scenario~\cite{choi2016millimeter}. This article presents a methodology and shows the steps to import real scenarios and realistic 3D objects for data creation in mobility scenarios that vary over time, and how the level of realism affects the computational cost. Simulated datasets do not replace, but complement, measurement data, which can improve and validate simulated data and statistical channel models as they become available.

The remainder of this paper is divided into four sections. The methodology for data generation is presented in Section~\ref{sec:methodology}. The integration of the scenarios for data creation is discussed in Section~\ref{sec:orchestrator}. It is discussed the computational cost of realistic models in the Vehicle-to-Infrastructure (V2I) data in Section~\ref{sec:results}, which is followed by the conclusions in Section~\ref{sec:conclusion}.

\section{Methodology for creating scenarios}\label{sec:methodology}

RT is considered a promising simulation strategy for 5G and provide very accurate results~\cite{fuschini_ray_2015,Rappaport14}. It is expected that increasing the level of confidence on RT simulators, the models should be as accurate as possible. For example, in outdoor scenarios, specific details are required, including accurate sizes, materials and format of buildings, a good distribution of the amount of vehicles and peoples, and advanced 3D models. It is important to emphasize that the realism of objects is proportional to the number of faces it contains, which results in more interactions, consequently greater computational cost. To to soften this consequence of the increase in the number of faces, optimized models were used in this article.

\begin{figure}[!hbt]
    \centering
    \includegraphics[width=0.4\linewidth,angle =-90]{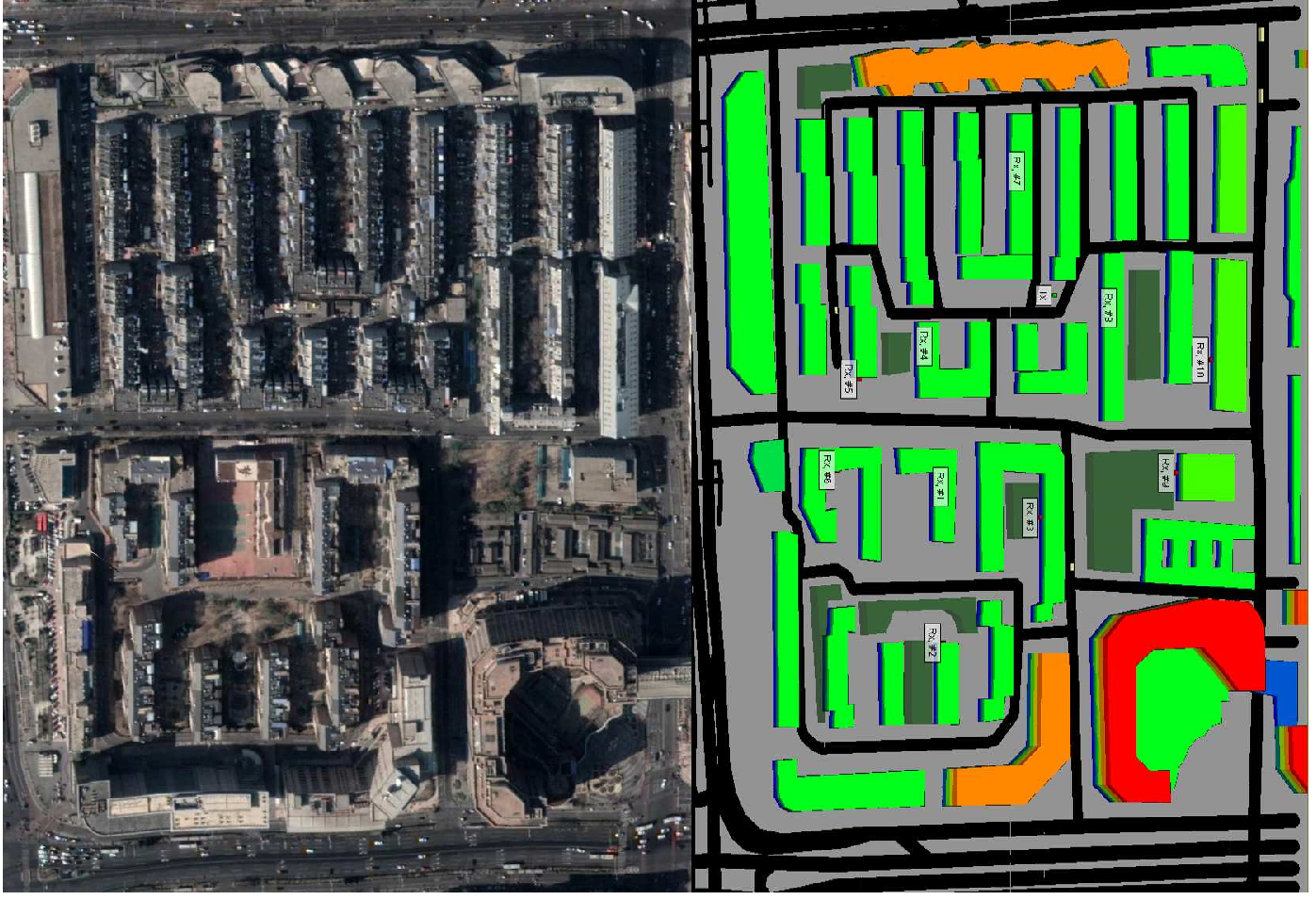}
    \caption{Selected study area: satellite image taken from Google Maps at left, and the DXF file imported to WI at right.}
    \label{fig:study_area}
\end{figure}

As a first step, it is necessary to choose a study area, where it is desired to analyze the propagation of the signal. To import a real scenario, the CadMapper was used, which is an online tool to transform data from public sources like OpenStreetMap, NASA and USGS in organized CAD files. In this article the study area chosen was Dongcheng, Beijing, area of China of approximately 0.3 square kilometers, in which only roads and buildings were used, as shown in Figure~\ref{fig:study_area}.

For realism, it is important to select well-mapped areas. The CAD file generated by this tool contains the 3D constructions, roadways, topography, vegetation and others. However, it is important to guarantee that all these features are in a format compatible with the RT simulator, for this work the  Wireless InSite (WI) was used. Assuming that the scenario was imported correctly, the next step is to set all the basic configuration in the RT simulator, such as the position of antennas, wave format, frequencies and others.

To obtain an improvement in the quality of data generated by the RT, 3D models that had formats closer to the real, were created using the Blender, which is a free software and open-source for creating 3D objects.

\begin{figure}[!hbt]
 	\centering
 	\includegraphics[width=0.45\linewidth]{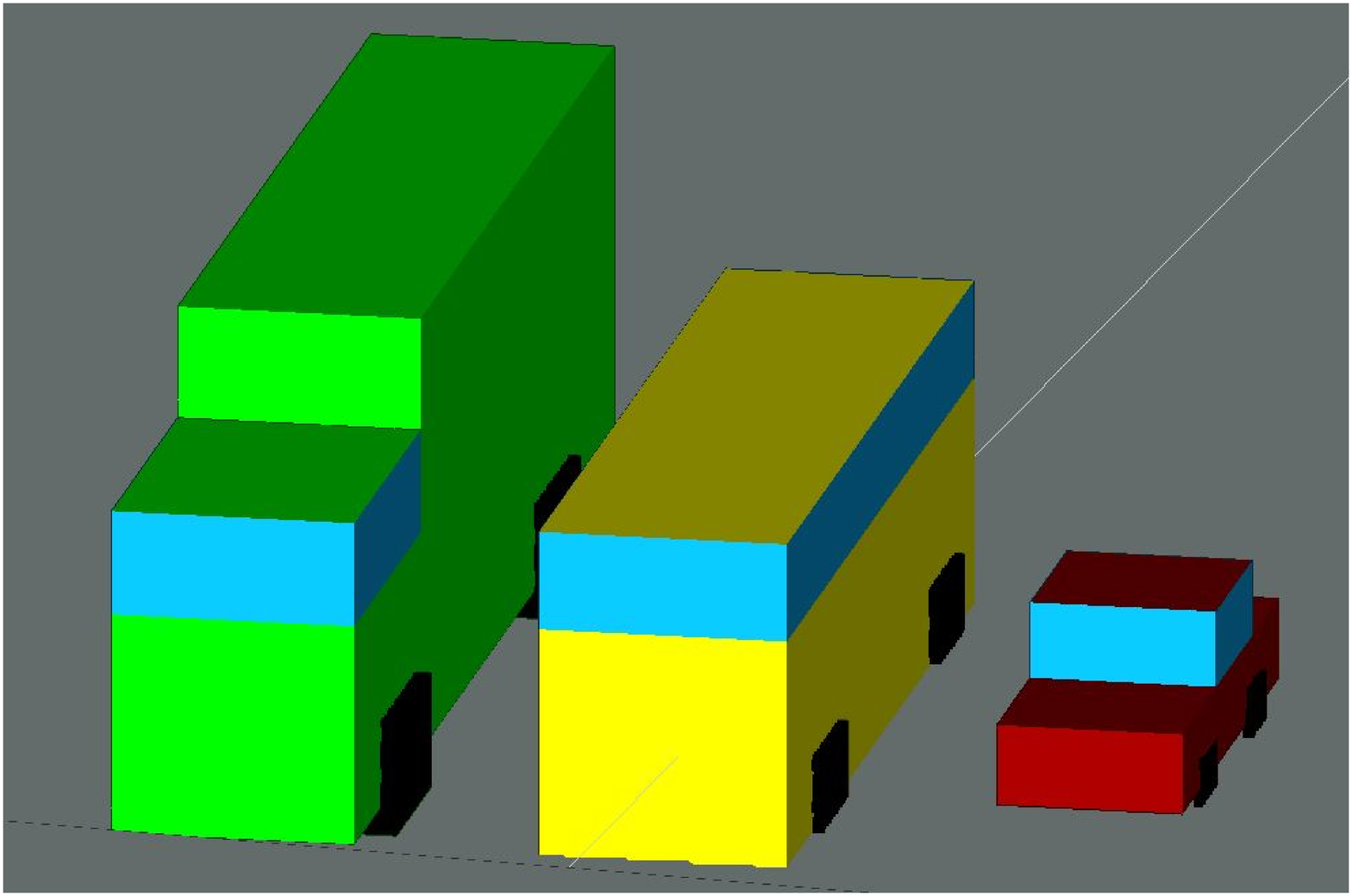}
 	\caption{Object models created using Blender.}
 	\label{fig:blender_models}
 \end{figure}

In order to create compatible environments with the daily life of cities, it was created car, bus, truck and pedestrian models. Due to increased computational cost when adding more realistic 3D objects with many faces, and WI causing lags when importing objects with curves, all objects were created in a minimalist way. However it is important to note that for a better representation of these objects in the RT software, faces must be created in places where the application of a specific material is desired, for example in car model, the region of the windshield will be associated the glass material while most of the vehicle will be made of metal, as can be seen in Figure~\ref{fig:blender_models}.
 
The mobility is inserted in the scenarios using Simulation of Urban Mobility (SUMO) software, which is an open-source traffic simulator that plays a key role in vehicle and pedestrian traffic flow generation so there is great diversity in RT simulation data~\cite{Behrisch11sumo}. To obtain the SUMO files for the simulation,  it is necessary to use the same coordinates data of the area selected in CadMapper. It is important to include all the necessary routes, and export it as OSM file. Based on this file, SUMO will generate other file which contains all roads of the area. After this, it is possible to set up vehicles and pedestrians flows, controlling speed, routes, acceleration and simultaneous generation.

\section{Orchestration between the simulators}\label{sec:orchestrator}

As mentioned before, in order to create realistic datasets it was adopted a simulation methodology that uses a traffic simulator and RT in mmWave communications. The base scenario is built in the WI, and the traffic simulation for this scenario is done using SUMO. After completing this step, a Python software~\cite{Rwisimulation} orchestrates the call of the simulators. Using the snap-shots of distinct moments from SUMO simulation, it is possible to obtain the position of the vehicles and pedestrians, thus positioning them as objects in the WI. The process is repeated during a time interval to produce the \enquote{mobile simulation}.

\section{Results} \label{sec:results}
\begin{figure}[!hbt]
	\centering
	\includegraphics[width=0.97\linewidth]{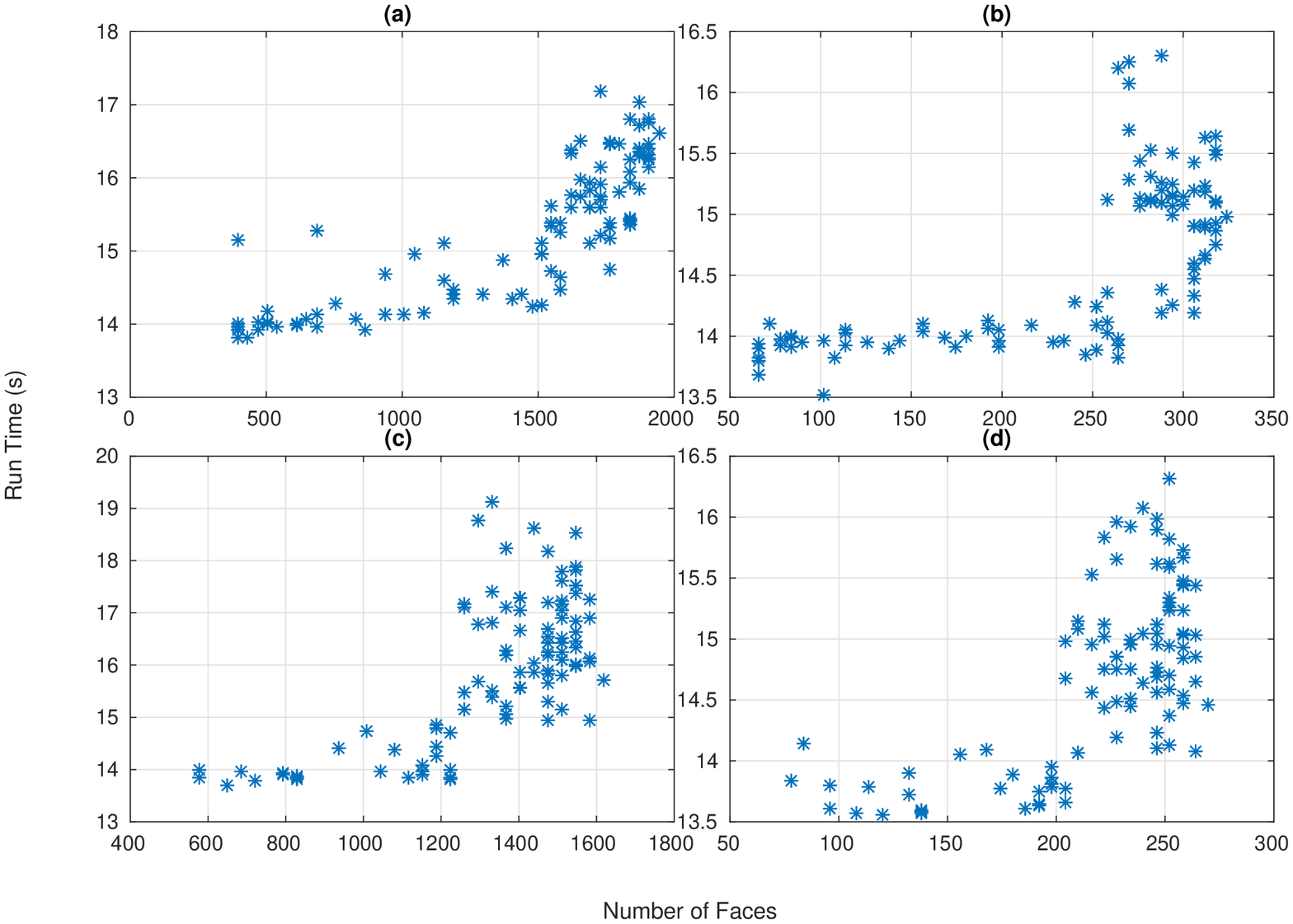}
	\caption{Comparison of the run time using the the two object models for two different receiver's configuration.(a) Fixed receivers, optimized models. (b) Fixed receivers, cube models. (c) Mobile receivers, optimized models. (d) Mobile receivers, cube models.}
	\label{fig:results_simulation}
\end{figure}

Some experiments were done to compare run time of the simulations with more detailed models and the cube models. Moreover, two type of receivers were analysed: fixed and mobile. The results of these experiments are illustrated in Figure~\ref{fig:results_simulation}. It can be inferred that using detailed models affects more the simulations which the receivers are mobile than the ones with fixed receivers. However, the overall mean values in both situations and using the different object models does not change drastically. 

\section{Conclusion}
This work briefly described the simulation methodology, the results show the 3D models increase the simulation time in only few seconds comparing the same moments on the simulation. This is because the 3D models are optimize to have less faces as possible, keeping the format of the object. Howsoever in research which request a large number of scenes, this few seconds will increase substantially the total time. The mobile simulations has a greater simulation time than the simulations with fixed receivers, because the mobility of the receivers increase the interactions each path can undergoes. Remains to analyze if this investment in more advanced models, make a real difference on the results of the channels analysis.

\label{sec:conclusion}

\bibliographystyle{IEEEtran}
\bibliography{sbrt}
\end{document}